\documentclass[aps,prl,twocolumn,showpacs,superscriptaddress,nofootinbib,floatfix]{revtex4-1}

\usepackage{graphicx}
\usepackage{amsmath}
\usepackage{amssymb}
\usepackage{color}
\usepackage{mathtools}
\usepackage[justification=RaggedRight]{caption}

\usepackage[breaklinks,colorlinks,citecolor=blue,linkcolor=red]{hyperref} 
\usepackage[all]{hypcap}

\newcommand{\aj}{AJ}

\renewcommand{\apj}{ApJ}
\newcommand{\apjl}{ApJL}

\renewcommand{\nat}{Nature}

\newcommand{\mnras}{MNRAS}

\renewcommand{\prd}{PRD}
\renewcommand{\prl}{PR:}

\newcommand\msun{\, \rm M_\odot}
\newcommand\BH{\mathrm{BH}}
\newcommand\GC{\mathrm{GC}}

\begin{document}

\title{Black hole mergers from an evolving population of globular clusters}
\author{Giacomo Fragione}
\affiliation{Racah Institute for Physics, The Hebrew University, Jerusalem 91904, Israel}

\author{Bence Kocsis}
\affiliation{Institute of Physics, E\"{o}tv\"{o}s University, P\'azm\'{a}ny P. s. 1/A, Budapest, 1117, Hungary}

\date{\today}

\begin{abstract} 
The high rate of black hole (BH) mergers detected by LIGO/Virgo opened questions on their astrophysical origin. One possibility is the dynamical channel, in which binary formation and hardening is catalyzed by dynamical encounters in globular clusters (GCs). Previous studies have shown that the BH merger rate from the present day GC density in the Universe is lower than the observed rate. In this \textit{Letter}, we study the BH merger rate by accounting for the first time for the evolution of GCs within their host galaxies. The mass in GCs was initially $\sim 8\times$ higher, which decreased to its present value due to evaporation and tidal disruption. Many BH binaries that were ejected long before their merger, originated in GCs that no longer exist. We find that the comoving merger rate in the dynamical channel from GCs varies between $18$ to $35\,{\rm Gpc}^{-3}\,{\rm yr}^{-1}$ between redshift $z=0.5$ to $2$, and the total rate is $1$, $5$, $24$ events per day within $z=0.5$, $1$, and $2$, respectively. The cosmic evolution and disruption of GCs systematically increases the present-day merger rate by a factor $\sim 2$ relative to isolated clusters. Gravitational wave detector networks offer an unique observational probe of the initial number of GC populations and their subsequent evolution across cosmic time. 
\end{abstract}

\maketitle

\paragraph{Introduction.---} 
The LIGO-Virgo Collaboration\footnote{https://www.ligo.caltech.edu/} has recently detected gravitational waves (GWs) from five binary black hole (BH) mergers, opening an entirely new window into high-energy physics \cite{abbott16,abbott17}. The astrophysical origin of these mergers is among the most puzzling open questions of our time. Possibilities include isolated binary evolution through a common envelope phase \cite{bel16b} or through chemically homogeneous evolution in short-period stellar binaries \citep{mand16,march16}, triple systems \cite{sil17,ant17,rodant18,arc18}, 
gas-assisted mergers \cite{bart17,sto17,tag18}, and
dynamically assembled binaries in dense stellar systems such as globular clusters (GCs) \cite{PZM00,rod15,askar17,baner18} or galactic nuclei \cite{okl09,ant12,ant16,ham18,hoang18}. 

In contrast to most other channels, the dynamical formation channel is theoretically well-understood, as it is determined by $N$-body gravitational interactions. In dense systems, chance multibody close encounters lead inexorably to the formation of BH binaries \cite{sig93}. Further scattering encounters decrease the BH binary separation. In at least 50\% of merging systems, the binary is ejected from the GC \cite{rod18,sam18a}, and the binary merges due to GW emission after several Gyr of its ejection \cite{rod16,askar17}. 

Previous studies of dynamically formed mergers in GCs have shown that the expected BH merger rate is $\sim5\,{\rm Gpc}^{-3}\,{\rm yr}^{-1}$ \cite{rod16,askar17,par17}, which is lower than the observed rate of $\mathcal{R} = 40$--$240\,{\rm Gpc}^{-3}\,{\rm yr}^{-1}$ reported by LIGO/Virgo, corresponding to a power-law mass function prior, and $\mathcal{R} = 12$--$65\,{\rm Gpc}^{-3}\,{\rm yr}^{-1}$ for a log-uniform mass function \cite{abb17}. The expected rates are sufficiently high that this contribution may be measured and distinguished from other channels statistically using the mass, spin, eccentricity, and redshift distribution \cite{olear16,zevin17,sam18a,sam18b,rod18}. In these studies the rates were estimated using the observed present-day GC density in the Universe and the GCs were evolved in isolation. 

In this \textit{Letter}, we point out the importance of including GC evolution within their host galaxies for studying the dynamically formed BH mergers. The initial mass in GCs is expected to have been a factor $\sim 8\times$ higher than today \citep{gne14,fag18}, since many GCs have evaporated and were tidally disrupted during interactions with the host galaxy. This expectation is confirmed by the observed radial and mass distribution of GCs in the Galaxy \cite{gne14}, and the observed high-energy emission from the Galactic bulge. The so-called Fermi-excess may have been produced by a population of millisecond pulsars that were formed in long-disrupted GCs \cite{bra15,fag18,fpb18,arc18a}. Does the increased initial GC mass increase the BH merger rate significantly? We determine the BH merger rate by accounting for the evolution and disruption of GCs in their host galaxies. We predict the redshift evolution of the merger rate, which may be measured with upcoming GW detectors. This may offer an observational probe to distinguish mergers of the dynamical GC channel from other astrophysical channels. If so, future GW measurements of the merger rate distribution offers an observational probe of the initial number of GC populations and their evolution. 

\paragraph{GC evolution.---} We follow Ref.~\cite{gne14} to evolve the initial GC population in their host galaxies using a semianalytical method. The GC formation rate is assumed to be a fixed fraction $f_{\mathrm{GC},i}=0.011$ of the total galactic star formation rate assuming that clusters formed at $z=3$ \citep{gne14,fag18}. We draw the initial mass of the clusters from a power-law distribution $dN/dM\propto M^{-2}$ from $10^4\ \mathrm{M}_{\odot} \leq M \leq 10^7\ \mathrm{M}_{\odot}$.

After their formation, GCs lose mass via three mechanisms, i.e. dynamical ejection of stars through two-body relaxation, removing stars by the galactic tidal field and stellar winds \cite{che90,hur00,kro01}. We evolve the cluster mass loss due to isolated evaporation due to two-body relaxation and stripping by the galactic tidal field according to
\begin{equation}
\frac{dM}{dt}=-\frac{M}{\min(t_{\mathrm{iso}},t_{\mathrm{tid}})}\ ,
\end{equation}
where \cite{gie08}
\begin{align}
t_{\mathrm{iso}}(M) &\approx 8.5\, M_5\ \mathrm{Gyr}\ ,\\
t_{\mathrm{tid}}(r,M) &\approx 2.07\, M_5^{2/3} \frac{r}{\mathrm{kpc}}\left(\frac{V_{\mathrm{c}}(r)}{200\mathrm{km}\ \mathrm{s}^{-1}}\right)^{-1}\ \mathrm{Gyr}
\end{align}
and where $M_5=M/10^5\ \msun$ and $t_{\mathrm{tid}}$ takes into account the strength of the local galactic field through the circular velocity $V_{\mathrm{c}}(r)$ at a distance $r$ from the galactic center. While in case of strong tidal field ($t_{\mathrm{tid}}<t_{\mathrm{iso}}$) the stars loss is dominated by the galactic tidal stripping, in the limit of a weak tidal field ($t_{\mathrm{tid}}>t_{\mathrm{iso}}$), the evaporation of stars is mostly controlled by internal dynamics. 

In proximity to the galactic center, the clusters may be torn apart due to the strong tidal forces. We assume that the cluster is disrupted when the average stellar density at half-mass radius falls below the mean galactic density
\begin{equation}
\rho_{\mathrm{h}}<\rho_*(r)=\frac{V_{\mathrm{c}}^2(r)}{2\pi G r^2}\ .
\label{eqn:dens}
\end{equation}
We adopt the average density at the half-mass radius
\begin{equation}
\rho_{\mathrm{h}}=10^3 \mathrm{M}_{\odot} \mathrm{pc}^{-3}\min\left[10^2,\max(1, 0.25\,M_5^2 )\right]\ .
\label{eqn:rhalfm}
\end{equation}
This equation limits $\rho_{\mathrm{h}}$ to $10^5\ \mathrm{M}_{\odot}$ pc$^{-3}$ in the most massive clusters, consistent with observations \cite{gne14}. We note that $\rho_*(r)$ in Eq. \eqref{eqn:dens} takes into account both the adopted field stellar mass, as well as the growing mass of the galactic bulge, that begins to build up as clusters are disrupted in the innermost galactic regions.

For what concerns the cluster orbit, following Ref.~\cite{gne14} for simplicity we evaluate the cluster at an instantaneous radial distance $r$ from the host galaxy center which represents the time-averaged radius\footnote{We include the effect of the deviation from circular orbit by including an eccentricity correction factor $f_e=0.5$ in the dynamical friction equation as in Ref.~\cite{gne14}.} of the true (probably eccentric) cluster orbit \cite{gne14}. We evolve the cluster orbits by evolving $r$ according to dynamical friction as\cite{bin08}
\begin{equation}
\frac{dr^2}{dt}=-\frac{r^2}{t_{\mathrm{df}}}\ ,
\label{eqn:dynf}
\end{equation}
where
\begin{equation}
t_{\mathrm{df}}(r,M)\approx 90 M_5^{-1}\left(\frac{r}{\mathrm{kpc}}\right)^2\left(\frac{V_{\mathrm{c}}(r)}{200\mathrm{km}\ \mathrm{s}^{-1}}\right)\ \mathrm{Gyr}\ .
\end{equation}

We describe the Milky Way's potential with a central $4 \times 10^6$ M$_{\odot}$ black hole, a Sersic profile with total mass $5\times 10^{10}$ M$_{\odot}$ and effective radius $4$ kpc, and a dark matter halo ($10^{12}$ M$_{\odot}$ with $r_{\mathrm{s}}=20$ kpc). Throughout the simulation, we constantly update the galactic mass distribution to include the stellar and gaseous debris from the disrupted clusters \cite{gne14}.

These initial conditions lead to the observed spatial and mass distribution of GCs surviving until $z=0$ in our Galaxy \cite{gne14,fag18}.

\paragraph{Rate of black hole mergers.---}

\begin{figure*}
\centering
\includegraphics[scale=0.55]{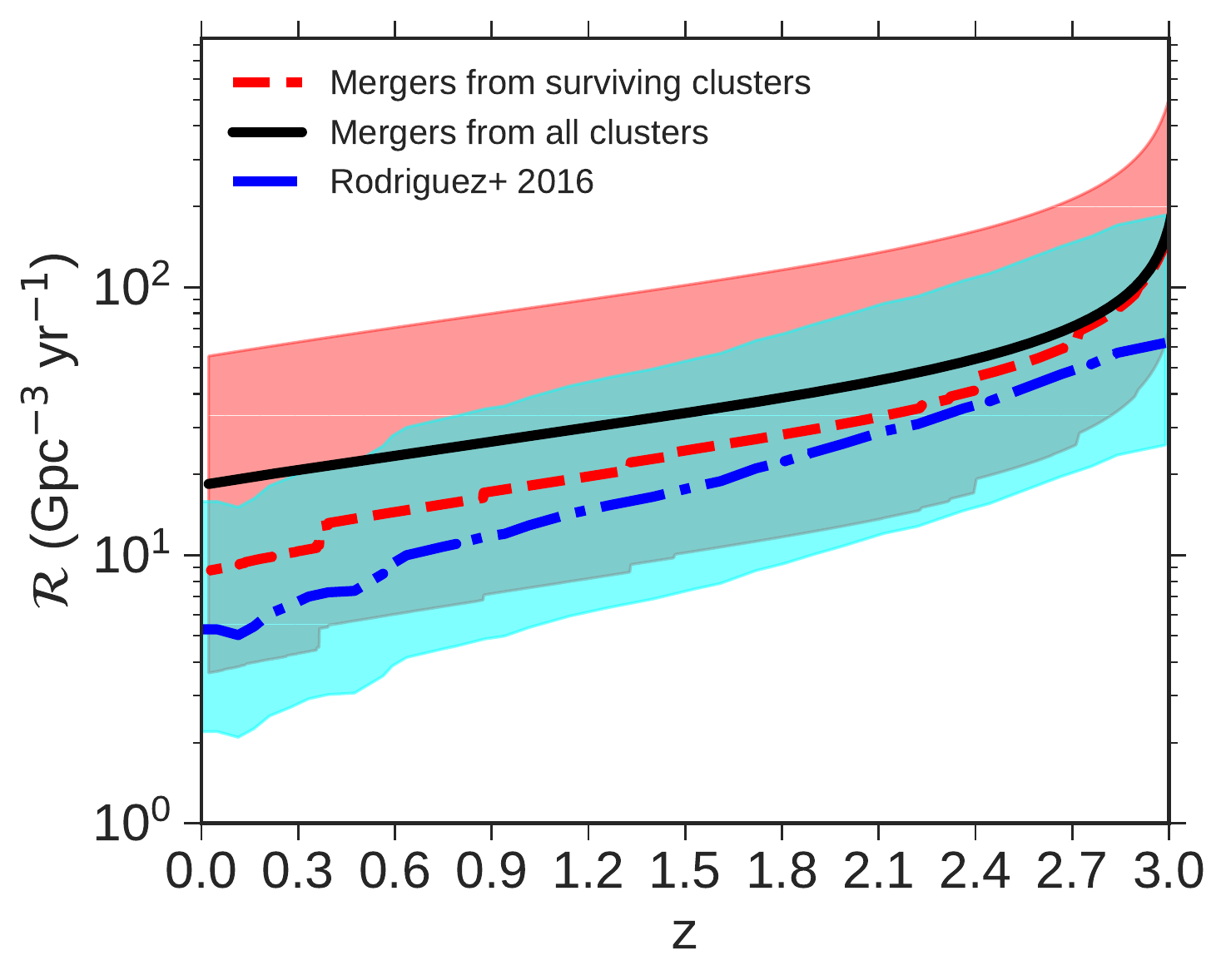}
\includegraphics[scale=0.55]{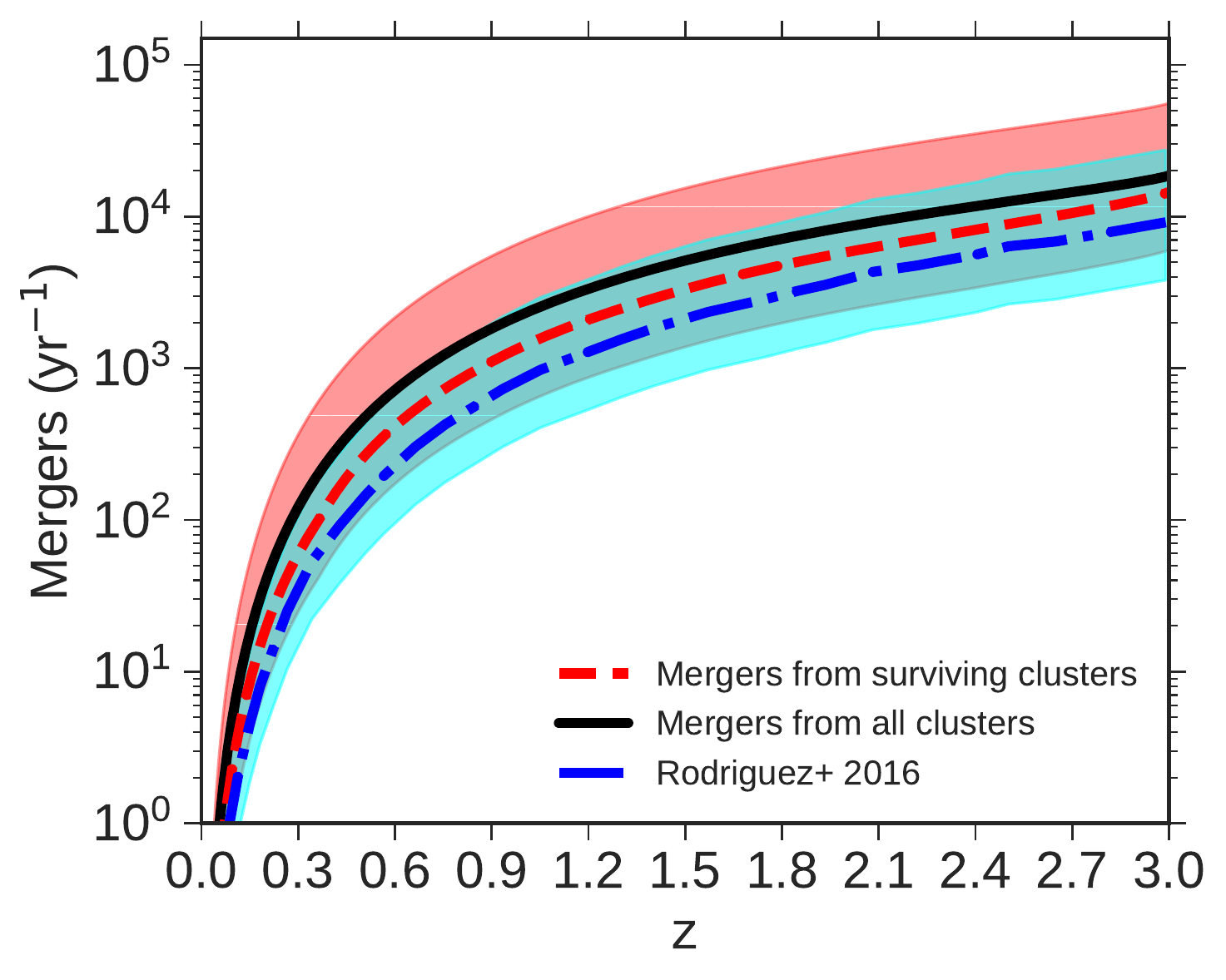}
\caption{The comoving BH merger rate density as a function of redshift $z$ (left) and the total number of sources that merge per unit observer time up to a maximum redshift $z$ (right). The black solid and red-dashed lines represent upper and lower limits on the expected rate from evolving GCs assuming respectively that merging binaries are all ejected before the cluster may be disrupted or that they merge within the cluster (see text). Blue dash-dotted line represents the result of Ref.~\cite{rod16} for isolated clusters. The merger rates are higher for evolving clusters that lost mass due to evaporation and tidal stripping, since they were initially more massive and more numerous to match the present day observed GC distribution. The shaded regions represent the model uncertainty by assuming $\rho_{\GC}$ in the range $0.32$--$2.31\,\mathrm{Mpc}^{-3}$ \cite{por00,rod15,rod16}, the lines represent $\rho_{\GC}=0.77\,\mathrm{Mpc}^{-3}$. The lower boundary of our model (red-shadowed region) is computed by scaling $\rho_{\GC}$ for the red-dashed curve, while the upper boundary is calculated similarly using the black-solid curve. The shaded blue region coresponds to Ref.~\cite{rod16}.}
\label{fig:rate}
\end{figure*}

We calculate the rate of mergers from the convolution of the 
$P_{{\rm ej},M}$ ejection probability at comoving time $t_{\rm ej}$ and the conditional merger probability among ejected binaries $P_{{\rm merg | ej},M}$ within inspiral time $\Delta t$ following ejection in a cluster of mass $M$ as
\begin{equation}
P_M(t) = \int_0^t P_{{\rm ej},M}(t_{\rm ej}) P_{{\rm merg | ej},M}(t-t_{\rm ej}) d t_{\rm ej}\,. 
\end{equation}
We fit the cumulative probability of BH-BH ejections from Morscher et al.~\cite{mor15} as
\begin{equation}\label{eqn:ejection}
C_{{\rm ej},M}(t) = f_{{\rm BH,ej},M} f_{{\rm bin,ej},M} \left(\frac{t}{t_{\rm H}}\right)^{0.4}\,,
\end{equation}
where $f_{{\rm BH,ej},M}\approx 0.5$ is the fraction of ejected BHs among all BHs in the cluster, and $f_{{\rm bin,ej},M}\approx 0.25$ is the fraction of BHs in ejected binaries relative to the number of all ejected BHs in a Hubble time $t_{\rm H}$\footnote{There may be significant ($\sim 30\%$) variations around these typical values depending on core radius \citep{mor15}.}.
We adopt the results of Monte-Carlo simulations by Rodriguez et al.~\cite{rod16} for the cumulative conditional probability distribution of the inspiral times of BH mergers $C_{{\rm merg | ej},M}(\Delta t)=\int_0^{\Delta t} P_{{\rm merg | ej},M}(t')dt'$  (see Figure 1 therein). We find that this function is fitted by
\begin{equation}
C_{M}(t) = c_1\,\mathrm{erf}[(\ln x)/c_2) ] +c_0\ ,
\label{eqn:C_M}
\end{equation}
where $c_1=0.497235$, $c_0=0.517967$, $c_2=5.69292$ and
\begin{equation}
x = \frac{t}{7.6\,\rm Gyr} \left(\frac{ M}{10^{6}\msun}\right)^4\,.
\label{eqn:x}
\end{equation}
This functional dependence follows from the conditions of binary ejection and the subsequent GW-driven evolution \cite{rod16}. Note that the timescale for a fixed fraction of BHs to merge within a GC is proportional to $M^{-4}$.

We compute the comoving merger rate density from the evolving population of globular clusters as a function of redshift by summing over the merger rate $\Gamma_{\BH}(z)$ over all GCs in a simulation in a Milky Way type galaxy, 
\begin{equation}
\mathcal{R}(z)=\frac{\rho_{\rm GC}}{N_{\GC}(0)} \sum_{i=1}^{N_{\GC}(z)}\,\Gamma_{\BH}(M_i,z)\ ,
\label{eqn:rategw}
\end{equation}
where $M_i$ is the initial mass of the $i$-th cluster, $N_{\GC}(z)$ is the number of GCs at redshift $z$ per Milky Way type galaxy taken from the simulation \cite{gne14} \footnote{In this estimate we neglect the fact that $N_{\rm GC}(z)$ may depend on galaxy type and galaxy mass.} and $\rho_{\GC}=0.77\,\mathrm{Mpc}^{-3}$ is the comoving number density of globular clusters \cite{rod15}.\footnote{This value of $\rho_{\GC}$ was used in Ref.~\cite{rod16} with which we compare our results. Previous work \cite{por00} found a factor 3.4 higher $\rho_{\GC}$.} In equation~\eqref{eqn:rategw}, the comoving rate of BH-BH mergers per GC of mass $M$ is obtained from the probability of BH-BH mergers as
\begin{equation}
\Gamma_{\mathrm{BH}}(M,t)=\frac{1}{2}f_{BH} M P_{M}(t)\ ,
\label{eqn:Gamma}
\end{equation}
where $f_{\BH}$ is the number of BHs per unit mass in a GC
\begin{equation}
f_{\BH}=\frac{\int^{m_{\max}}_{m_{\rm crit}} f_{\mathrm{IMF}}(m) dm}{\int^{m_{\max}}_{m_{\min}} m f_{\mathrm{IMF}}(m) dm}
\end{equation}
and $f_{\mathrm{IMF}}(m)$ is the stellar initial mass function \cite{kro01}
\begin{equation}
f_{\mathrm{IMF}}(m)=k
\begin{cases}
(m/0.5\msun)^{-1.3}& \text{$m_{\min}\le m\leq 0.5\msun$},\\
(m/0.5\msun)^{-2.3}& \text{$0.5\msun\le m\leq m_{max}$},
\end{cases}
\label{eqn:imf}
\end{equation}
where $m_{\rm min}=0.08\msun$, $m_{\rm max}=150\msun$, and $m_{\rm crit}=20\msun$ is the critical mass above which BHs form. In reality there may be a transitional mass range allowing either NSs or BHs to form below $20\msun$ \cite{Sukhbold2018} and the effective $m_{\rm crit}$ may also depend on  metallicity \cite{gimap18} or the host cluster \cite{cho18}. A lower effective value, e.g. $m_{\rm crit}=17\msun$, would produce $25\%$ more BHs and imply a higher merger rate, and vice versa for a higher $m_{\rm crit}$. Further, the details of the BH ejections due to kick-velocities at birth \cite{repnel15}, and the relative retention fraction, may affect the compactness of the host GC \cite{mack08}. These factors, as well as primordial binaries, may have some effect on the results presented in this paper and deserve further study. We convert time from the initial redshift $z_{\rm in}=3$ to $z$ to use in Eqs.~\eqref{eqn:rategw} and \eqref{eqn:Gamma} using the cosmological relation \cite{eis97}.

\begin{figure*}
\centering
\begin{minipage}{18cm}
\hspace{-0.5cm}
\includegraphics[scale=0.55]{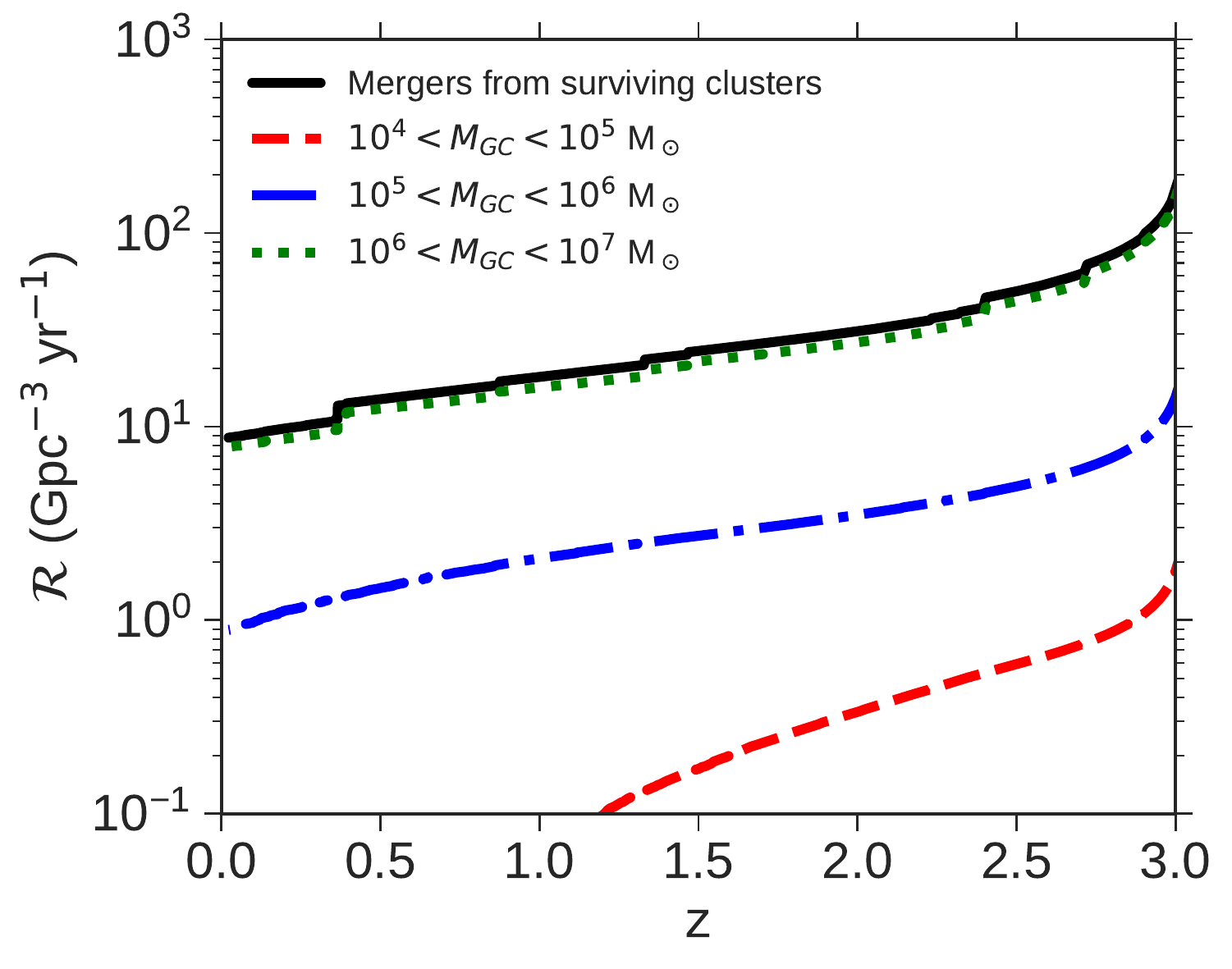}
\includegraphics[scale=0.55]{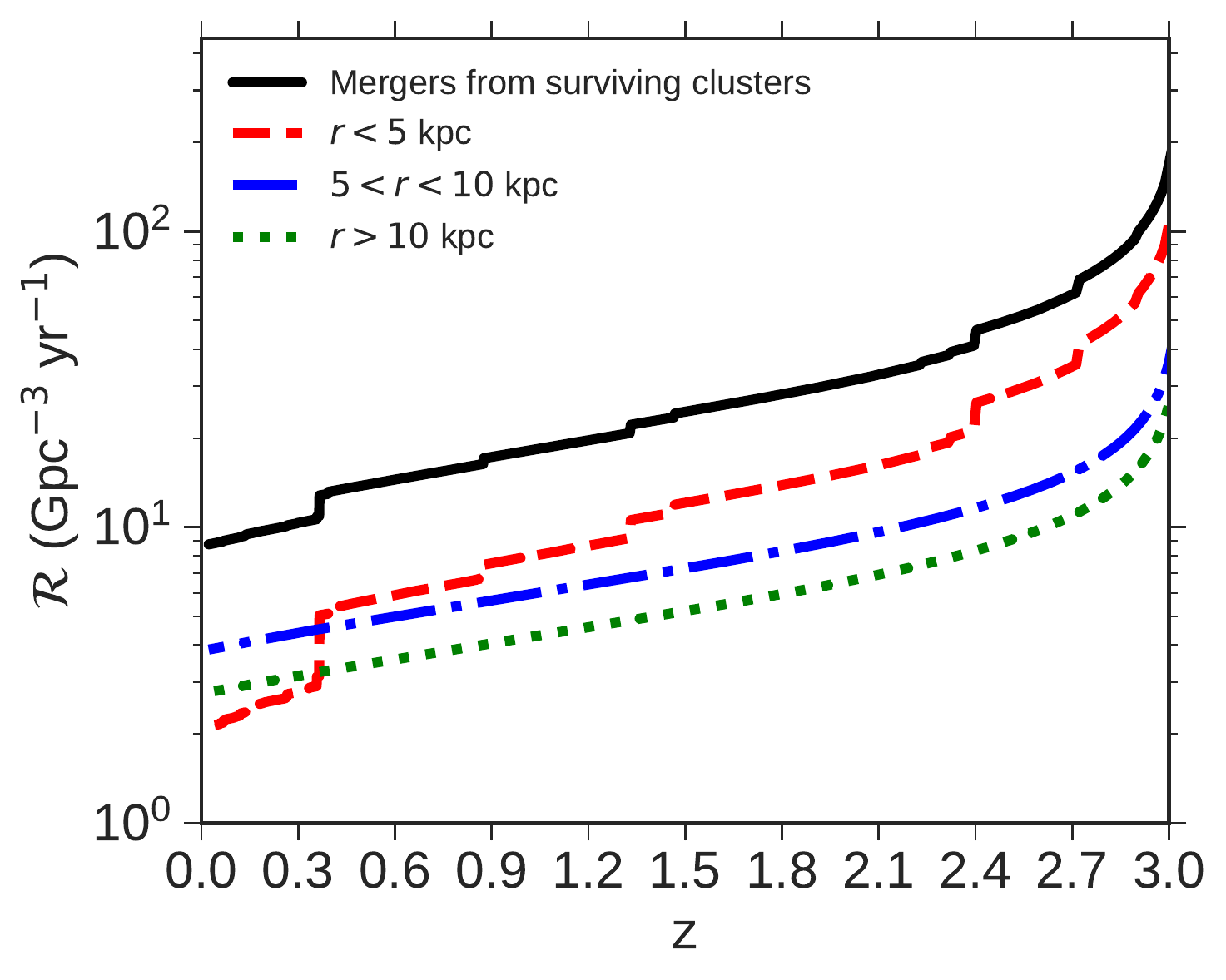}
\end{minipage}
\caption{The comoving BH merger rate density as a function of redshift $z$ subdivided in the different contributions by GCs of different initial masses (left) and different initial position in the host galaxy (right).}
\label{fig:rate_mr}
\end{figure*}

Dynamically formed BH binaries merge both within the host GC and far outside of it after ejection. Recently it was pointed out that accounting for GW losses during the dynamical evolution of GCs increases the fraction of mergers within the host GC significantly to $50\%$ \cite{rod18,sam18a}. This affects the merger rate given by equation~\eqref{eqn:rategw}. For mergers that happen inside the GCs at redshift $z_{\rm merger}$, only the existing GCs ($N_{\rm GC}(z_{\rm merger})$) at that redshift contribute to the rate. However, for BHs that merge after their ejection from the host cluster (Eq.~\ref{eqn:ejection}), all clusters ($N_{\rm GC}(z_{\rm ejection})$) must be included at the point of ejection, $z_{\rm ejection}$. The latter assumption implies a higher $N_{\rm GC}$, thus producing a higher rate of mergers. For a robust estimate, we calculate the BH merger rate as a function of redshift from the probability distribution of mergers (Eq.~\ref{eqn:rategw}) in two limiting cases: either extending the sum over all GCs\footnote{this corresponds to $N_{\rm GC}(z=3)$ in the adopted model \cite{gne14}} or up to $N_{\rm GC}(z)$, respectively. The true merger rate must be between these bounds.

Apart from the rate density, it is useful to calculate the total merger rate in the Universe within redshift $z$:
\begin{equation}
\mathcal{C}(z)=\int_0^z \frac{dV_c}{dz'} \frac{R(z')}{1+z'}\ dz'\ ,
\end{equation}
where $dV_c/dz'$ is the comoving volume at redshift $z'$ and the $1/(1+z')$ factor accounts for the observed redshifted time at Earth compared to the source's comoving time.

\paragraph{Results.---}
Figure \ref{fig:rate} shows the BH merger rate from the evolving population of GCs in the Universe as a function of redshift. The left panel shows the comoving BH merger rate density $\mathcal{R}$. The red dashed and black solid lines provide the lower and upper limits assuming mergers inside surviving clusters at redshift $z$ and ejected mergers from all clusters as specified above. The black line represents the case where most ejections leading to mergers happen early, well before possible GC disruptions. The blue curve shows the previous result \citep{rod16} for isolated GCs for comparison. The difference between our upper and lower bounds are relatively small (factor $\lesssim 2$ at $z\sim 0$ and smaller at higher $z$), since the rate is dominated by the initially more massive clusters that survive until today. The merger rate is a factor of $\sim 2$ larger for the evolving GC population than for the case of isolated clusters computed in Ref.~\cite{rod16} at $z\lesssim 1$. This increased merger rate is a result of the originally more massive clusters in comparison to isolated clusters due to tidal stripping (see Eqs.~\ref{eqn:C_M}--\ref{eqn:x}). The right panel of Fig. \ref{fig:rate} shows $\mathcal{C}(z)$, the total rate of mergers within redshift $z$. The shaded regions represent model uncertainty by assuming $\rho_{\GC}$ in the range $0.32$--$2.31\,\mathrm{Mpc}^{-3}$ \cite{por00,rod15,rod16}.

Figure \ref{fig:rate_mr} presents the comoving BH merger rate density as a function of redshift for different intervals of initial GC masses (left panel) and position in the galaxy (right panel). Although the most abundant, low-mass clusters ($10^4 \msun<M_{GC}< 10^5 \msun$) contribute to a negligible fraction to the total rate as they are inefficient at merging (see Eqs.~\ref{eqn:C_M}--\ref{eqn:x}) and because they dissolve. The largest contribution near the formation epoch of GCs comes from the most massive population ($10^6 \msun<M_{GC}< 10^7 \msun$), whose contribution is $\sim 6$-$7$ times that of $10^5 \msun<M_{GC}< 10^6 \msun$. In terms of the radial distribution of mergers (right panel), the rate is dominated by clusters in the inner galaxy ($r<5$ kpc), apart from small redshifts ($z\lesssim 0.3$). Roughly 25\% of the rate comes from GCs in the outer halo ($r>10$ kpc).

\paragraph{Discussion---}
\label{sect:conclusions}
In this \textit{Letter}, we have determined the BH merger rate from dynamically formed binaries produced in GCs that coevolve with their host galaxies in the Universe. At redshift $z=0$, we have found a rate $\sim 4$--$60\,{\rm Gpc}^{-3}\,{\rm yr}^{-1}$, within the uncertainties of our model. We found that the expected merger rate ranges between $\mathcal{R} \sim 18\,{\rm Gpc}^{-3}\,{\rm yr}^{-1}$ to $\sim 35\,{\rm Gpc}^{-3}\,{\rm yr}^{-1}$ for redshift between $z=0.5$ to $2$, and the total rate is $1$, $5$, $24$ events per day within $z=0.5$, $1$, and $2$, respectively. This corresponds to a factor $\sim 3$ to a $\sim 2$ higher rate from $z=0.5$ to $z=2$ with respect to the case neglecting the evolution of GCs in their host galaxies. If a significant fraction of mergers from GCs is from ejected binaries, the rate at low redshift $z<0.1$ is $\sim 10\,{\rm Gpc}^{-3}\,{\rm yr}^{-1}$, a factor of $\sim$2 higher than previous estimates. For comparison, the current observational limit on the BH merger rate in the local Universe is $\mathcal{R} = 12$--$65\,{\rm Gpc}^{-3}\,{\rm yr}^{-1}$, assuming a log-uniform mass distribution, and $\mathcal{R} = 40$--$240\,{\rm Gpc}^{-3}\,{\rm yr}^{-1}$ for a power-law BH mass distribution with $dN/dm \propto m^{-2.35}$.

This result highlights the need for more detailed simulations of GCs tracking their evolution with their host galaxies. Our results were derived by scaling to Milky-Way type hosts. Future work is needed to study the discrepancy between the rates of evolving GC populations and isolated GCs for different host galaxies, whose GC population correlates with the dark matter masses \cite{for16}. Furthermore, the possible presence of an intermediate-mass black hole (IMBH) may significantly change the evolution of GCs and the distribution of merger rates \cite{fragk18,frlgk18}.

At design sensitivity, LIGO--Virgo is expected to observe BH mergers up to $z\sim 1$ \cite{Voyager}. Assuming that the LIGO--Virgo observational completeness is 1--10\% (i.e. the fraction of all mergers that LIGO--Virgo detects within this volume), our results suggest that one detection per day to $\sim 1$ per week is expected from the dynamical channel. Recently, Fishbach et al.~\cite{fish18} suggested that with $\sim 100-300$ LIGO--Virgo detections it will be possible to distinguish among different models of the merger rate evolution within the coming 2-5 years. Indeed, the redshift evolution of merger rate from the dynamical channel shown in Figure~\ref{fig:rate} is distinct from other formation channels as it increases from $z=0$ until the epoch of globular cluster formation with a particular shape as shown (see \cite{bel16b,fish18} and references therein, for the redshift evolution for other channels). The cosmic evolution and disruption of GCs increases the present-day merger rates by a factor $\sim 2$ in comparison to isolated clusters (see Fig.~\ref{fig:rate}). The redshift evolution of the merger rates carries information on the cosmic history of GCs. Thus, measuring the redshift evolution of the rates will represent an observation probe of GC formation, their initial numbers in the Universe and their evolution across cosmic time. With a sufficiently large sample of mergers the relative contribution of intergalactic GCs \cite{cal14,Mackey16} may be distinguished from GCs evolving in galaxies. Future instruments such as the Voyager, Einstein Telescope, or Cosmic Explorer will make this endeavor more feasible \cite{punt10,Voyager}.

We conclude that GW detectors have the potential to provide a view on the evolution of faint GCs which are practically invisible to electromagnetic observatories. This may have far reaching implications in the theory of galaxy formation, possibly leading to the understanding of the theory of GC formation and the origin of the empirical correlations between the number of GCs, their host SMBH, and dark matter halos \cite{harris16}.

\bigskip
\begin{acknowledgments}
We thank Oleg Gnedin and Carl Rodriguez for useful discussions and comments and the anonymous referees for helpful suggestions. GF is supported by the Foreign Postdoctoral Fellowship Program of the Israel Academy of Sciences and Humanities. GF also acknowledges support from an Arskin postdoctoral fellowship and Lady Davis Fellowship Trust at the Hebrew University of Jerusalem. This project has received funding from the European Research Council (ERC) under the European Union's Horizon 2020 research and innovation programme ERC-2014-STG under grant agreement No 638435 (GalNUC) and from the Hungarian National Research, Development, and Innovation Office grant NKFIH KH-125675 (to B.K.).
\end{acknowledgments}

\bibliography{refs}

\end{document}